\documentclass[12pt]{article}

\usepackage{titling}

\usepackage{scicite}

\usepackage{times}

\usepackage{graphicx}

\newenvironment{sciabstract}{\begin{quote}\bf}{\end{quote}}

\title{Structure of Native Two-dimensional Oxides on III--Nitride Surfaces}

\author
{J.~Houston Dycus,$^{1}$ Kelsey J.~Mirrielees,$^{1}$ Everett D.~Grimley,$^{1}$\\ Ronny Kirste,$^{2}$ Seiji Mita,$^{2}$ Zlatko Sitar,$^{1,2}$ Ramon Collazo,$^{1,2}$\\ Douglas L.~Irving,$^{1}$ James M.~LeBeau$^{1}$\\
\\
\normalsize{$^{1}$Department of Materials Science and Engineering, North Carolina State University,}\\
\normalsize{Raleigh, North Carolina 27695, USA}\\
\normalsize{$^{2}$Adroit Materials, Inc., 2054 Kildaire Farm Rd., Suite 205,} \\
\normalsize{ Cary, North Carolina 27518, USA}\\
\\
\normalsize{$^\ast$To whom correspondence should be addressed; E-mail: jmlebeau@ncsu.edu.}
}

\newcommand{\alumina}{Al$_2$O$_3$}
\newcommand{\gallia}{Ga$_2$O$_3$}

\begin{document}

% Two-dimensional structures have primarily been discovered amongst previously known structures, e.g. graphene from graphite \cite{graphene, mos2, etc}. Interest in these systems has arisen from often unique physical properties made possible by inherent quantum confinement \cite{2DGaN}. Further, their  properties are often critically influenced by their structure and defects \cite{2DGaN, Graphene, etc}.

 %For example, GaN and AlN surfaces are known to form oxides, but the details of these structures have largely been inferred.


\baselineskip24pt


\maketitle

\begin{sciabstract}
When pristine material surfaces are exposed to air, highly reactive broken bonds can promote  the formation of surface oxides with structures and properties differing greatly from bulk. Determination of the oxide structure, however, is often elusive through the use of indirect diffraction methods or techniques that probe only the outer most layer. As a result, surface oxides forming on widely used materials, such as group III-nitrides, have not been unambiguously resolved, even though critical properties can depend sensitively on their presence.  In this work, aberration corrected scanning transmission electron microscopy reveals directly, and with depth dependence, the structure of native two--dimensional oxides that form on AlN and GaN surfaces.  Through atomic resolution imaging and spectroscopy, we show that the oxide layers are comprised of tetrahedra--octahedra cation--oxygen units, similar to bulk  $\theta$--Al$_2$O$_3$ and $\beta$--Ga$_2$O$_3$.  By applying density functional theory, we show that the observed structures are more stable than previously proposed surface oxide models. We place the impact of these observations in the context of key III-nitride growth, device issues, and the recent discovery of two-dimnesional nitrides.% Beyond these immediate implications, these two--dimensional oxides also present new opportunities to control device properties.

\end{sciabstract}

% For example, oxidized surfaces can vastly improve the reactivity of metal catalysts \cite{Lundgren:2002a} and are key to electronic properties of buried interfaces. 

%on the improved development of a vast range of applications ranging from lighting to water purification as well as demonstrate that two--dimensional oxides form natively on the surface of III--nitrides.

%The reconstruction results in an inversion boundary formed at the surface mediating a polar inversion in subsequent growth.

%\end{abstract}

%While III--nitrides are a proven and mature material for both opto- and power- electronic devices, much remains unknown at their surfaces

%In particular, group III--nitrides are promising for semiconductor applications such as blue and UV light emitting diodes (LEDs) \cite{Mueller-Mach:2002aa,Akasaki:1994aa}. As with other optoelectronic materials, the properties are intimately linked to the atomic structure and deviations from the parent structure such as strain or defects will alter properties such as band gap \cite{Mattila:1997aa,Stampfl:2000aa,Yan:2009aa}. 

%either by extrinsic doping or from surface states.

Group III--nitrides, such as GaN and AlN, are the bedrock of modern solid--state lighting.  Further, they are of particular interest for high power devices because of their strong polarization fields that can confine carriers to heterointerfaces supplied via extrinsic doping or  surface states \cite{vandewalle:2014np}. At the surface of the III-nitrides, however, dangling bonds are reactive to their environment \cite{Zhu:2012aa, shibata2008direct,Xu:2016ab}, and lead to the formation of surface oxides.  Importantly, the surface structure has been proposed to play an important role in controlling the electronic properties of buried device structures.  For example, in power electronic devices,  surface compensation can have a dramatic influence on the mobility of the two-dimensional electron gas (2DEG) that forms at AlGaN/GaN heterointerfaces  \cite{Shur:1996aa,Li:1997aa,Jogai:2003aa,Koley:2005aa}.  

%This degradation has been tackled empirically using passivization to improve device reliability.  Without details of how the surface structure impacts properties, however, it is difficult to make widespread use of the technology. 

While further advancements require complete understanding of the native surface oxides,  much remains unknown about their structure \cite{Liao:1993aa,Bermudez:1996aa,Dalmau:2007aa,Li:2006aa,Rice:2010aa}. Methods such as X-ray photoelectron or Auger electron spectroscopies and electron diffraction \cite{Marks2010:aa} provide key insights, but are an indirect probe. While scanning tunneling microscopy can probe the surface atomic and electronic structure, the outer most valence electrons are primarily probed. As a result, surface oxide structural models have thus largely relied on presumed atomic configurations paired with first principles density functional theory (DFT) calculations to estimate relative stability of each model \cite{Dong2006,Miao:2010aa}. As a result, there are a number of competing surface oxide models that can be difficult or impossible to determine without direct surface \textit{and} sub-surface information from experiment.

%Such requirements are not limited to devices exploiting 2DEG but in general to any device based on such polar materials as III-nitrides \cite{citeneeded}.  

%Of the proposed surface oxides, the surface structure was modified to satisfy electron counting or stoichiometric conditions. For the electron counting rule, it was found that O needed to replace N at the surface sites \cite{Miao:2010aa}, consistent with previous results from XPS. 

%Elucidating the nature of III--Nitride surfaces can thus provide the opportunity to manipulate electronic properties of these critical materials.

Here, we report the direct observation of native two--dimensional oxides that form on III--nitride surfaces. Using aberration corrected scanning transmission electron microscopy (STEM) imaging and spectroscopy, the structure of these two-dimensional oxides is directly determined for c--plane AlN and GaN surfaces.  The observed oxides differ considerably from bulk structures, but with bonding configurations consistent with the corresponding group III oxide. Furthermore, these oxides are found to be more energetically stable than previous surface oxide models over a wide range of chemical environments. Finally, the structures are discussed in the context of material growth and properties. 
    
%Nitride based semiconductors serve as the basis for current and future optoelectronics with applications ranging from water purification to efficient lighting. As with other optoelectronic materials, the properties are intimately linked to the atomic structure.  

 \begin{figure}[ht!]
 \begin{center}
 \includegraphics[width=3.1in]{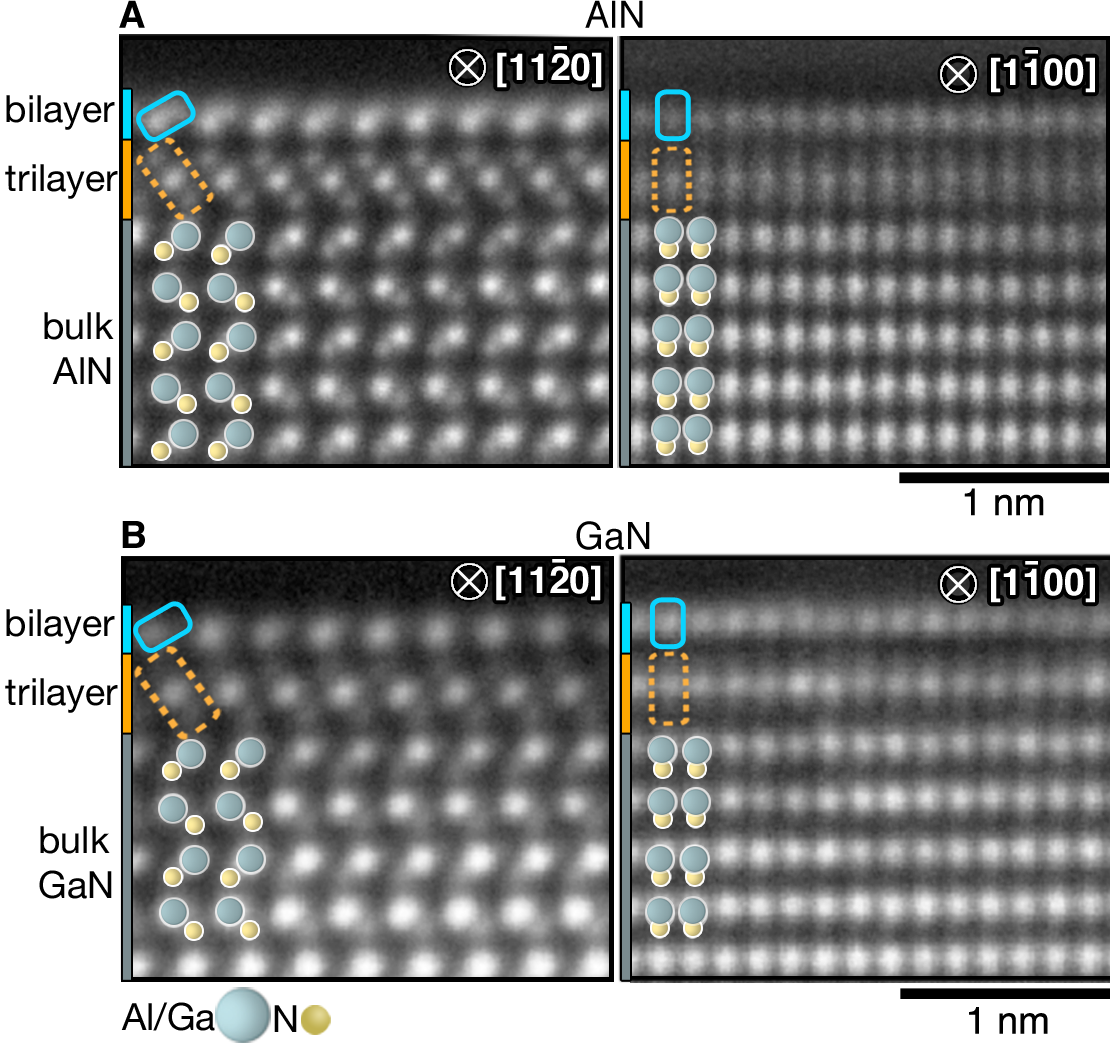}
 \caption{Native two-dimensional oxides observed on the c--plane of (A) AlN and (B) GaN  via ADF STEM.  Perpendicular view directions are presented (left/right). The dashed (orange) and solid (blue) boxes highlight the trilayers and bilayers that comprise the oxide structure.}
 \label{fig:1v4}
 \end{center}
 \end{figure}

 %shows orientated along the [11$\bar{2}$0] (a) and [1$\bar{1}$00] (b) directions
 
%After a short etch with HF, 

Direct, atomic resolution annular dark-field (ADF) STEM reveals the formation of a surface layer on  c--plane AlN and GaN as shown in Figure \ref{fig:1v4}. For both materials, the surface is comprised of a two--dimensional layer that forms across the c-plane sample surface.  The layer is distinguished by a trilayer (dashed boxed regions) and an inverted bilayer (solid boxes) in Figures \ref{fig:1v4}A and B.  The outer surface is consistent with prior oxide formation studies using indirect methods and STM, but those methods could not  resolve the sub-surface information \cite{Dong2006}. Furthermore, the layered structure appears passivated as further growth of the oxide does not occur over time.  Also, as discussed in the Supplementary Information, various approaches to the TEM sample preparation were attempted to rule out unintentional modification of the surface. Regardless of the preparation approach, the observed surface structure remained the same.

% A single two-dimensional layer was always observed

\begin{figure}[ht!]
 \begin{center}
 \includegraphics[width=3.13in]{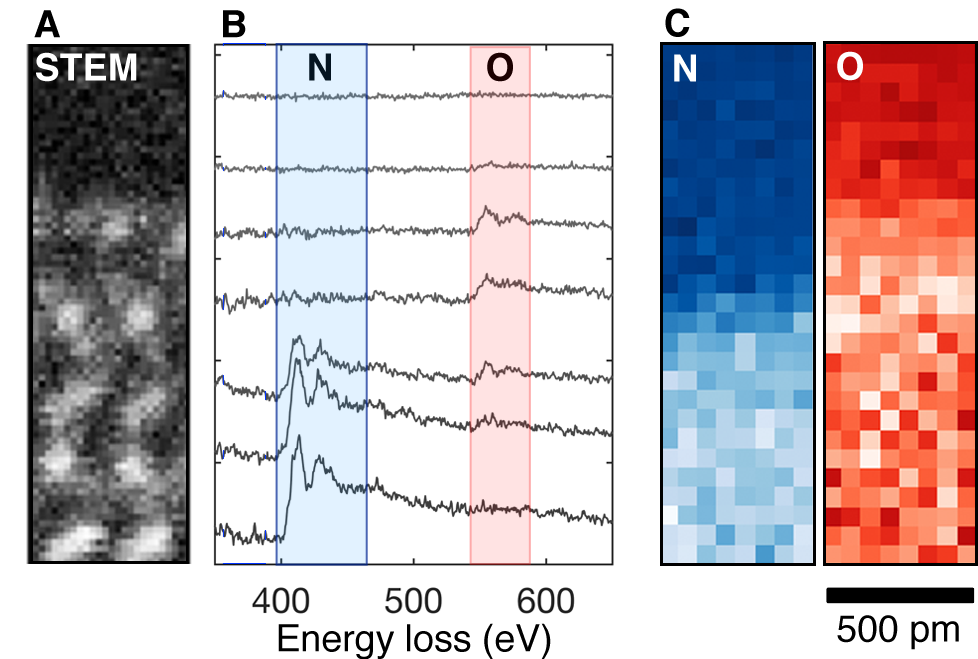}
 \caption{(A) ADF STEM and corresponding (B) integrated EEL spectra rom across the c--plane surface. (C) Elemental distribution of the surface determined by background subtracting and integrating the spectral signal over the range indicated in (B).}
 \label{fig:EELS}
 \end{center}
 \end{figure}

While atom column positions can be directly determined from the ADF STEM images, the atomic species can be difficult or impossible to identify without additional information \cite{Houston-Dycus:2013aa}. For unambiguous elemental analysis, we turn to electron energy loss spectroscopy (EELS), where an abrupt transition from the nitride to a two-dimensional oxide layer is seen in Figure \ref{fig:EELS}. Furthermore, the oxygen signal extends into the nitride beyond the surface oxide, which may indicate oxygen--nitrogen intermixing, delocalization of the electron probe, and/or oxide formation on the top/bottom of the TEM sample. Note that EELS of the GaN surface exhibits the same distribution of anions, as shown in Figure \ref{fig:GaNeels}. 

%Furthermore, other surface terminations did not exhibit the formation of the two-dimensional oxide, as shown in Figure S\ref{fig:mface} for the m-plane AlN surface. 

% , Al--N = 192 pm or Ga--N = 199 pm  \cite{Costales:2002aa}

\begin{figure*}[ht!]
 \begin{center}
 \includegraphics[width=6.5in]{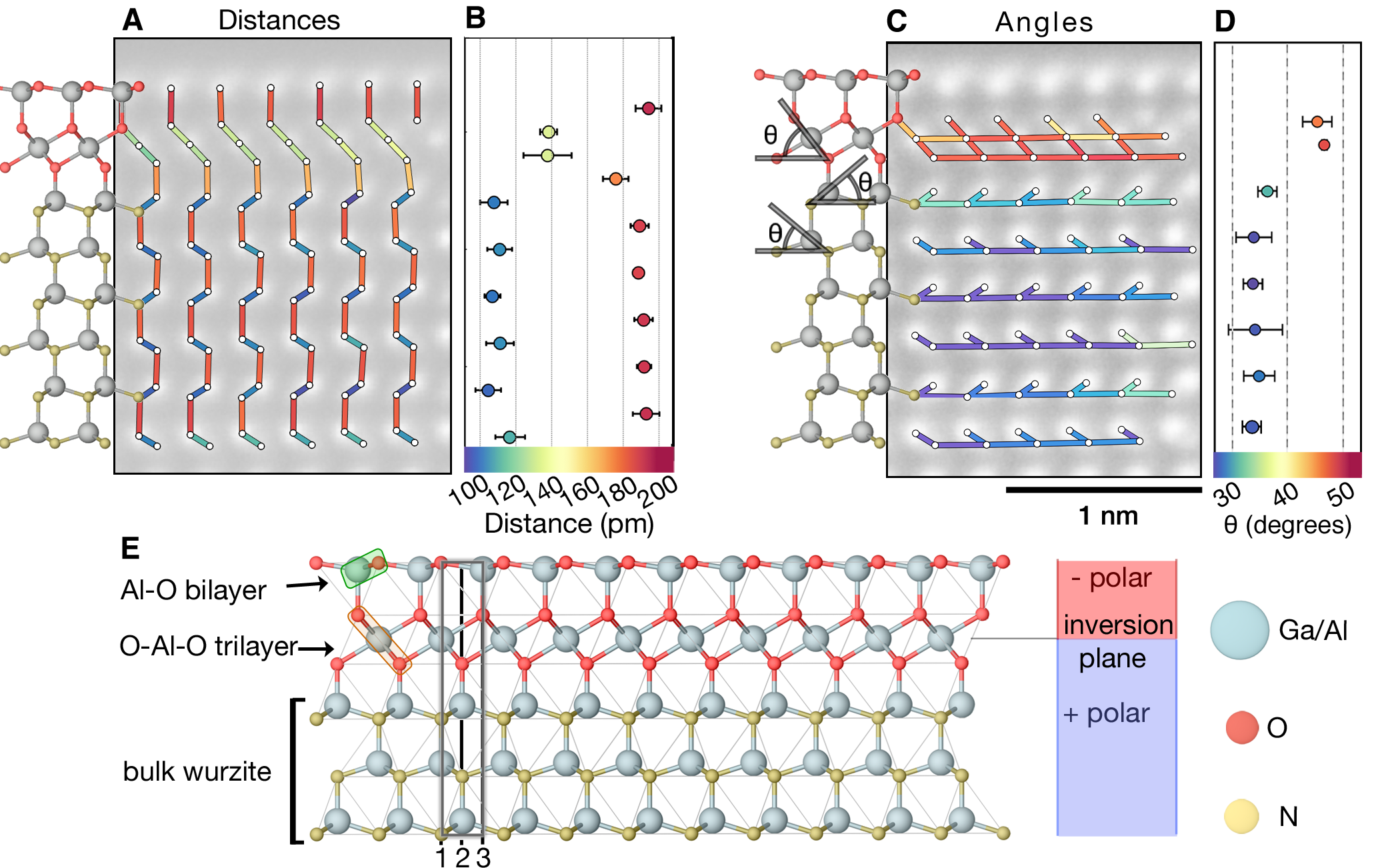}
 \caption{ (A) Distances measured within the two-dimensional oxide and corresponding (B) layer averages.  (C) The projected bond angle, $\theta$, measured as depicted in the schematic and corresponding (D) layer averages. The color of each measurement corresponds to the relevant colorbar provided at the bottom (C,D). (E) Model of the AlN surface oxide constructed using the distance, angle, and chemical information from EELS. }
 \label{fig:distances}
 \end{center}
 \end{figure*}

Distance measurements from the perpendicular view directions in Figure \ref{fig:1v4} are used to construct a three dimensional model of the oxides.  Through use of revolving STEM \cite{Sang:2014aa}, accurate and precise \cite{Dycus:2015aa} atom column distance measurements for AlN are included in Figure \ref{fig:distances}.  The projected Al--N distances measured in the AlN substrate agree with those (expected) from bulk: 109 pm (107 pm) and 191 pm (192 pm). At the  nitride/oxide interface, the first Al--O bond length is 176 pm. This length is consistent with the shorter bonds in bulk Al oxide, Al--O 175 pm  \cite{Jones:1968aa}, than the nitrides. In the middle of the trilayer, the projected Al--O distance is 138 pm (Figure \ref{fig:distances}A,B).  A comparison to GaN distance measurements is provided in Figure \ref{fig:alnandgandist}.  In addition, in--plane distances between adjacent Al/Ga and N/O columns gradually decreases, as shown in Figure \ref{fig:inplane}. 

To probe bond angles, the angle between neighboring anions on (0001) planes and the nearest group III atom column are measured, as shown for  AlN in Figure \ref{fig:distances} (C,D). The average angles are 34$^\circ$ and 48$^\circ$ for the AlN bulk and oxide, respectively.  These are consistent the projected angles from wurtzite and Al$_2$O$_3$, which are 35$^\circ$ and 52$^\circ$, respectively.

%The final oxide layer expands out of plane to 194 pm. 

%For the case of GaN, the large difference in atomic number decreases the ability to finely separate the Ga and N neighboring columns, but comparisons between the AlN and GaN surface oxides can be drawn by measuring the distances between the metal atoms (Al, Ga) for each structure. Supplemental Figure \ref{fig:alnandgandist} shows the distance between metal atoms along the [0001] direction. Within the bulk the measured Ga-Ga projected distance is 12 pm longer than the Al-Al distance in AlN, consistent with the difference in bonding between structures. In the oxide, an increase in the group III atom column separation is observed. Although each increase, the ratio of increased distance between III--III sites for GaN to AlN decreases to 10 pm and 3 pm for the final III site in the bulk to the III site in the triplet and the trilayer III position to the surface bilayer III site respectively. The result may indicate that a relative difference in the bonding occurs for each surface layer.

 %Combining the measurements above with geometry based on imaging from a perpendicular direction, $\left[1\bar{1}00\right]$  Figure \ref{fig:1v4}, a 3D structural model is constructed.
 
 The model constructed from the structural measurements provides insights into atom coordination of the oxides.  First, for both Ga and Al oxides, the last cation layer at the oxide/nitride interface remains tetrahedrally coordinated with N at the base and O at the apex of the tetrahedron. Second, highly distorted cation--O octahedra are observed at the central oxide  layer. This configuration appears encouraged by the tetrahedral coordination at the c--plane surface.  While this mixture of coordination is not present in the stable $\alpha$-\alumina{} phase \cite{Zhou:1991aa,Lee:1997aa}, the  two-dimensional structure observed here is structurally similar to $\theta$-\alumina{} and $\beta$--\gallia{}, which both exhibit coexisting tetrahedral and octahedral cation coordination.  It is also noted in Figure \ref{fig:distances}E that there is an inversion of polarity across the mid-plane of the oxide trilayer based on this configuration, which suggests a role of these surface oxides in compensating the strong, built--in polarization field of the wurtzite nitride. 
 
 %An illustration of the surface reconstruction for each observed orientation is shown below the experimental images with a horizontal line indicating where the inversion plane occurs and the abrupt transition from cation--polar to anion--polar.

 %The $\theta$-\alumina{} and $\beta$-\gallia{} type
 
 % DFT details moved up, should discuss model and use trends of distances to validate.

  %According to DFT results, the trilayer is more energetically stable.
  
\begin{figure}[ht!]{}
\begin{center}
\includegraphics[width=3.3in]{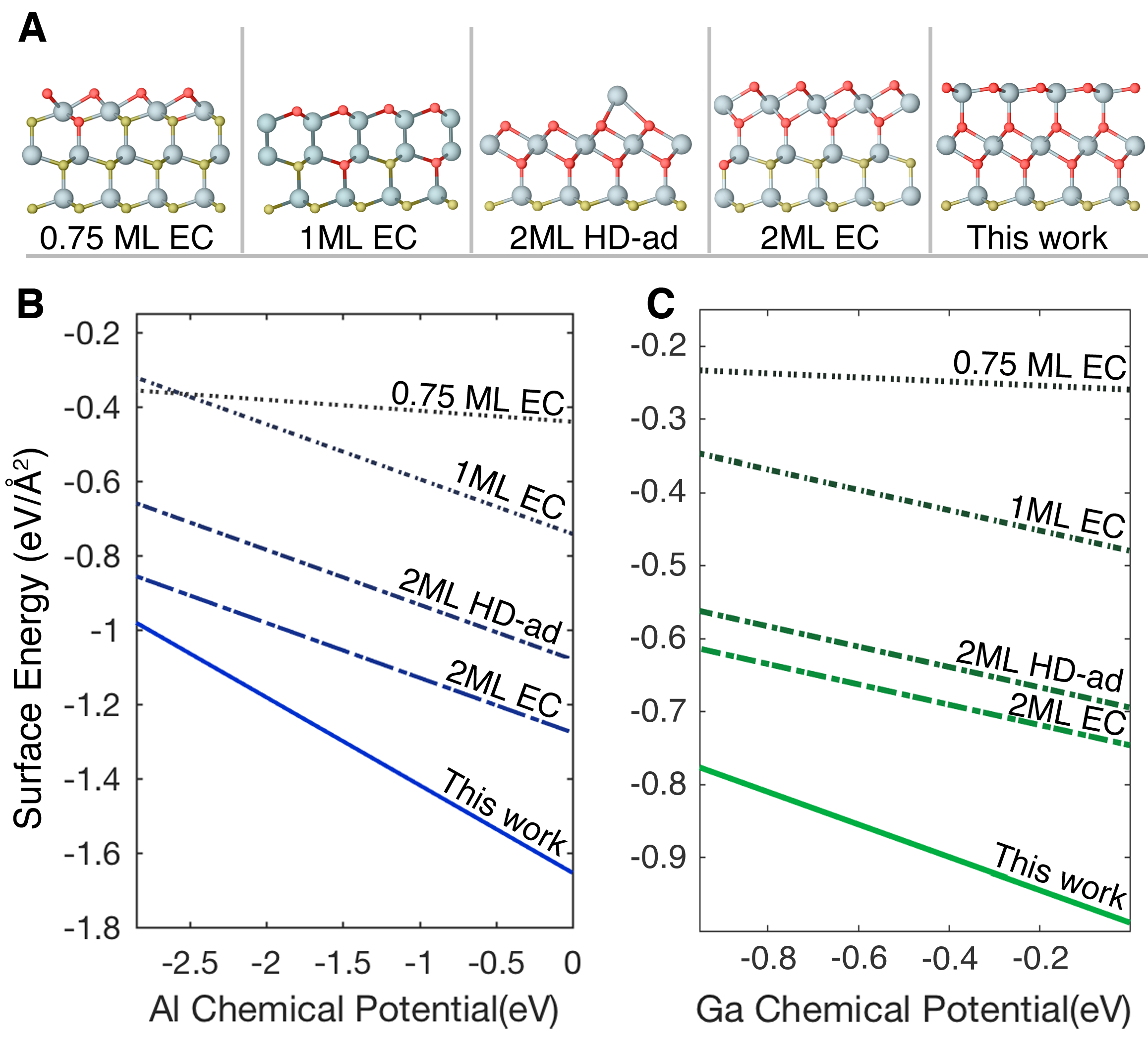}
\caption{Surface formation energies for previously proposed oxide structures and the structure observed in this work. (A) The surface oxide models with ML, HD, and ad indicated multilayer, high density, and adsorbed, respectively. Surface energies as a function of (B) Al , and (C) Ga chemical potentials with an oxygen chemical potential of -0.3 eV.}
\label{fig:DFT}
\end{center}
\end{figure}

Though the oxidized surfaces of GaN and AlN have been studied and modeled for some time, this direct observation of the surface with depth resolution reveals a new structure that differs from those previously investigated. To understand the energetic stability of this structure, we compare it with previously proposed models \cite{Dong2006,Miao:2010aa} using DFT.  In those works, it was found that the most energetically favorable structures consist of either an octahedrally coordinated O--III--O trilayer or a tetrahedrally coordinated III--O bilayer which can be seen at the top of Figure \ref{fig:DFT}. An ideal version of the observed structure maintains the bulk oxide 2/3 cation-to-anion ratio.  This is achieved by combining aspects of both trilayer and bilayer models, which was not previously considered.

%This structure can be made to satisfy electron counting rules through the introduction of cation vacancies. 

The surface formation energy for each surface configuration is calculated via Equation \ref{surface_energy}. This surface energy is taken relative to a clean, smooth, and step-free cleaved nitride surface.  Negative surface formation energies indicate that the oxidized configurations are favorable, for the particular set of conditions, relative to the unreconstructed and ideally flat surface. Absolute surface energies taken relative to the bulk are a challenge in this direction due to the lack of inversion symmetry along $\left[0001\right]$. Nevertheless, the relative energies provide insight into the favorability of one configuration as compared to another. The surface energy as a function of chemical potential is given by:
% Surface energy equation
% \begin{equation}
% \label{surface_energy}
% {\Delta}E^f(\mu_{Al/Ga}) = E^{\mathrm{tot}}_{\mathrm{slab}} -  E^{\mathrm{tot}}_{\mathrm{ref}} - n_{Al/Ga}\mu_{Al/Ga}  - n_N\mu_N - n_O\mu_O
% \end{equation}

%More general equation for surface energy
\begin{equation} 
\label{surface_energy}
E^f = \frac{1}{A}\left[E^{\mathrm{tot}}_{\mathrm{slab}} - E^{\mathrm{tot}}_{\mathrm{ref}} - \sum_in_i\mu_i\right]
\end{equation}

In this equation, $E^{tot}_{slab}$ is the total energy of the slab model containing the reconstruction, $E^{tot}_{ref}$ is the energy of the eight bilayer reference, $A$ is the area of the surface, $n$ is the number of atoms added to (positive) or removed from (negative) the surface structure with atoms being exchanged with a chemical reservoir described by the chemical potential $\mu_i$, where $i$ is the element (Al, Ga, N, or O) being exchanged. The total chemical potential, $\mu_i$, is $\mu_i= \mu_i^o +\Delta\mu_i$, where $\mu_i^o$ is the reference chemical potential taken at 0 K and $\Delta\mu_i$ is the change in the chemical potential from that reference.
 
The calculated surface energies, $E^f$, as a function of the Al and Ga chemical potential for the observed structure and previously proposed models, are presented in Figure \ref{fig:DFT}. Based on an environment  of air with an oxygen partial pressure of 0.2 atm at room temperature (298 K), $\Delta\mu_O$ is fixed at -0.3 eV \cite{mallard2000nist}. Of the previously reported structures, those consisting of an O--Al--O trilayer were shown to be more energetically stable. For the observed structure, $E^f$ is lower than all the previous models across the range of Al and Ga relevant chemical potentials. 

%The lower $E^f$ for the oxide on AlN versus GaN indicates that oxidation of AlN surface is energetically more preferred than that of the GaN surface for the environmental conditions, consistent with experiment \cite{oxideStability}. 

%These results strongly support the conclusion that the observed structure is native and not formed by TEM sample preparation or by interaction with the high energy electrons during STEM imaging. Additionally, the stability of this surface is such that amorphous regions at the edge of ion milled samples without subsequent chemical etching are observed to reconstruct towards the trilayer+bilayer structure over the course of a few minutes electron irradiation.

%Considering realistic oxidation conditions, the chemical potential ranges between -0.32 eV at T = 300 K and -1.40 eV at T = 1200 K under a pressure of 1 atm. Within this range, the proposed structure is shown more energetically favorable to form. At room temperature and atmospheric conditions, the observed structure has the lowest $\Delta$E$^f$, indicating this is the stable native oxide phase and not resulting from imaging of the electron beam. Further, the same structure was observed when an image was acquired without prior electron beam irradiation. 

Using the relaxed DFT model based on the observed structure, STEM image simulations, Figure \ref{fig:supplSimulation}, are in excellent agreement with the experiment.  While the STEM images from experiment shows that the oxide exhibits a slightly lower overall intensity compared to simulations, this can be due either incomplete coverage of the oxide on the sample surface, differences in static and thermal atomic displacement factors, unoccupied cation sites, or a combination thereof. To further validate the surface oxide model, the bond lengths and angles are compared between experiment and DFT. The bond length measurements are shown in Table \ref{tab:STEMvsDFT}. The trends for each bond in and between the nitride and oxide are in excellent agreement. Quantitatively, the  bond lengths exhibit at most $\sim$5\% error. The largest deviation occurs at the oxide/nitride interface. Further, the agreement extends to the  bond angles, matching experiment and theory to within 0.5$^\circ$.

%cation column intensity as compared to the bulk cation column intensity. 

 %We performed an initial analysis using DFT of structures containing cation vacancies that also satisfy electron counting rules. The surface formation energies were found to be on the same order or slightly lower than the oxide stoichiometric structure discussed above. Nevertheless, these layers  have more significant deviations in bond lengths and angles as compared to experimental measurements. This is due, in part, to the artificial periodicity of the vacancies in the supercells.  It is also important to note that the measurement standard deviation in experiment is greater for the oxide layer. The increased standard deviation further suggests additional disorder in the surface oxide that is not readily incorporated into the DFT calculation.

%The maximum deviation in bond lengths is 10 pm (5.4 \%), however, both distances are shorter than the bulk Al--N distance.

Given the structure of the stable 2D oxides, there are a number of consequences. First, inversion domain boundaries are often found in thin film nitrides \cite{Mohn:2016ab}, and the observed oxides offer a mechanism for their creation.  Notably, the oxides inherently reverse polarity through the transition from octahedral to tetrahedral bonding as discussed above. The inversion of polarity in the outer most layer of the oxide would seed N-polar nitride growth. Supporting this hypothesis, oxygen has been found at the initiation site of inverted domains, forming AlN$_{1-x}$O$_x$ at elevated growth temperatures \cite{Mohn:2016aa}.  Although the 2D oxide and the AlN$_{1-x}$O$_x$ by Mohn et~al.~differ in structure, they consist of similar octahedral and tetrahedral bonding units. Differences between the two structures may result from reaction of the 2D oxide seed layer during subsequent high temperature growth.

The structure of the observed oxide also explains the results from a recent study growing ZnO grown epitaxially on GaN. In that case, the surface of the GaN was intentionally oxidized before ZnO growth, where the ZnO polarity was then inverted \cite{Ullah:2016a}. The authors of that work hypothesized that a monolayer oxide was responsible for the polarity inversion and high quality of growth.  The structural models provided here therefore provide critical insight into reducing defects as well as improved heteroepitaxial oxide--nitride thin film structures. 

These 2D oxides appear similar to the recently reported formation of a 2D form of GaN  \cite{2DGaN}.  The structure, stabilized by encapsulation with graphene, also  exhibits octahedral -- tetrahedral bonding in a trilayer -- bilayer configuration. In both cases, the polarity is inverted across the central trilayer, which could aid in charge compensation. As such, this structural configuration may be a hallmark of Ga- and Al- based 2D oxides and nitrides. Further, these similarities motivate future directions of research in the development of a mechanistic understanding of such 2D structures on polar surfaces and demonstrates that even though the nitrides have been studied for decades  new discoveries remain.

%Here, EELS confirms the predominance of oxygen in our 2D structure.

%It is also possible that the previously observed structure may have been the same oxide observed here and resulted from oxygen contamination during or after growth.

%Finally, this observation leads to important consequences for the formation of 2DEG's in nitride based HEMTs. It has been emperically found that passivating the surface, for example with Si$_3$N$_4$ is required to improve device performance.

In summary, by combining experiment and theory, we have the directly solved the structure of two-dimensional oxides forming natively on AlN and GaN surfaces. These oxide structures provide key observations to explain  the formation of inversion domains and the origin of surface states that significantly influence the performance of III-nitride based devices. Further, these models offer direct evidence to model electronic surface states within the bulk band structure. Because of the relative ease of forming these structures, we also propose that such a platform may provide opportunities for exploring the properties and electronic behavior of two dimensional oxides not previously considered.

\bibliography{refs}
\bibliographystyle{Science}

%% classes.   This should be given within the "acknowledgement"
%\begin{acknowledgement}

\section*{Acknowledgements}

JML, JHD, and EDG gratefully acknowledge support for this research from the Air Force Office of Scientific Research (Grant No.~FA9550-14-1-0182).
DLI and KJM acknowledge financial support from the NSF under grant DMR-1151568.
JHD and EDG acknowledge support by the National Science Foundation Graduate Research Fellowship (DGE-1252376).  R.K., S.M., R.C., and Z.S.~sincerely thank support from NSF (ECCS-1508854, ECCS-1610992, DMR-1508191, ECCS-1653383), and the Army Research Office (W911NF-15-2-0068, W911NF-16-C-0101) for funding. This work was performed in part at the Analytical Instrumentation Facility (AIF) at North Carolina State University, which is supported by the State of North Carolina and the National Science Foundation (ECCS-1542015).  AIF is a member of the North Carolina Research Triangle Nanotechnology Network (RTNN), a site in the National Nanotechnology Coordinated Infrastructure (NNCI).  

%\end{acknowledgement}

\section*{Author Contributions}
%I don't know if there is a correct way to write this section but please amend any of the names if it incorrect. 
J.H.D and J.M.L designed the experiments and carried out the STEM imaging and subsequent image analysis. K.J.M. and D.L.I performed the DFT calculations and analysis.  E.D.G performed the multislice STEM image simulations of the structural model.  R.K., S.M., R.C., and Z.S.  performed the sample growth.   J.M.L., D.L.I., R.C., and Z.S.~supervised the research.  All authors contributed to discussions and preparation of the manuscript.  

\clearpage
\section*{Supplementary materials}
Materials and Methods\\
Figures S1 to S4\\
Table S1 \\
References \textit{(30-39)}

\clearpage

\setcounter{page}{1}
% \begin{center}
% 
% \end{center}

\title{Supplementary Materials for: Structure of Native Two-dimensional Oxides on III--Nitride Surfaces}

\maketitle

\renewcommand{\thefigure}{S\arabic{figure}}
\setcounter{figure}{0}

\section*{Materials \& Methods}

%For sapphire substrates, multiple steps were required to account for the large lattice mismatch between the substrate and film.  First, the sapphire substrates were annealed at 1100 $^\circ$C under pressure near 10$^{-6}$ Torr, followed by H$_2$ annealing and nitridation \cite{Mita:2008aa}.  Next, a 10 nm low-temperature AlN buffer layer was deposited at 640 $^\circ$C and annealed at 1050 $^\circ$C to transition from N--polar to Al--polar in the film.  An intermediate high temperature AlN layer was then deposited using a growth condition that promoted 3D growth to reduce the dislocation density by bending dislocations towards free surfaces.  The resulting dislocation density was on the order of 10$^{10}$ cm$^{-2}$ and growth of thicker films without cracking.  Finally, the growth conditions were changed to promote step flow growth, minimizing roughness for the sequential steps during AlN growth.  For imaging a single crystal of AlN, substrates were obtained from single crystal AlN boules grown by physical vapor transport \cite{Zhuang:2006ab,Zhuang:2006aa,Lu:2009aa} from HexaTech.

%\hl{GaN growth description needed.}

\subsection*{Experiment}
Thin film AlN and GaN samples were prepared by MOCVD growth following methods described in Ref. \cite{dalmau2011growth,mita2008influence}. Thin film cross sections were prepared for TEM using the methods described in Refs.~\cite{Dycus:2017aa} and \cite{Voyles:2003aa}. Subsequent to mechanical polishing, samples were prepared with and without Ar ion milling to determine the impact of the milling process.  For this step, a Fischione 1050 TEM mill was for final thinning at 1.0 keV and 0.4 keV for 10 minutes and 20 minutes respectively.  It is noted that the same structure is observed for samples that were prepared with and without ion milling.  Similarly, the samples were imaged with and without plasma cleaning, and the same surface structure was observed under both conditions.  To remove any damaged surface layer during sample preparation, a dilute 5\% HF solution was used to etch the TEM samples for one minute.  It is noted that the same structure is observed for each variation of the preparation step. 

A FEI Titan G2 60--300 kV TEM/STEM was operated at 200 keV for spectroscopy and 300 keV for imaging. For AlN, the annular dark-field collection range was 34-208 mrad at 200 keV and 30-185 mrad at 300 keV.  The probe--forming convergence semi-angle was 19.6 mrad in each case. For GaN, the annular dark-field collection range was 22 mrad and 134 mrad at 200 keV and 19 mrad and 118 mrad at 300 keV.  To minimize beam damage, the probe current was kept below 30 pA. Sample thicknesses were approximately 5 nm as determined by position averaged convergent beam electron diffraction \cite{LeBeau:2010ab}

To improve signal to noise and remove sample drift distortion, the RevSTEM method was used \cite{Sang:2014aa}. Image series were acquired with at least 20 1024$\times$1024 pixel frames at a pixel dwell time of 3 $\mu$s/pixel with a 90 $^\circ$ rotation between each frame.    Atom column positions were fit to a two-dimensional Gaussian distribution with sub--pixel precision and indexed into a matrix   \cite{Sang:2014ab}.  The pixel size was calibrated as described in Ref.~\cite{Dycus:2015aa}.  EELS maps were acquired using a Gatan Enfinium spectrometer.  The O and N maps were extracted individually after fitting and subtracting the spectral background via fitting to a power law.  Image simulations were carried out using the multislice method \cite{Kirkland:2010dq} with parameters taken from experiment. Thermal scattering was included with the frozen phonon approach and thermal displacements approximated by their bulk values.

\subsection*{Simulation and calculation details}

DFT plane wave pseudopotential calculations, as implemented in the Vienna ab--initio Simulation Package (VASP)\cite{kresse1993ab,PhysRevB.49.14251,kresse1996efficiency,PhysRevB.54.11169},  were used to determine surface formation energies. The gradient corrected Perdew-Burke-Enzerhof (PBE)\cite{PhysRevLett.78.1396} functional was used to handle exchange and correlation contributions to the total energy.  Projector augmented pseudopotentials for Al, Ga, N,and O contained 3, 13, 5, and 6 valence electrons, respectively. A kinetic energy cutoff of 520 eV was used for all structures. Each oxide was oriented along a cleaved and unreconstructed (0001) plane of the wurtzite AlN/GaN structure that consisted of eight bilayers. A 20 {\AA} vacuum padding was also included to minimize surface-surface interactions between both sides of the slab. Exposed N atoms at bottom of the slab were passivated with partially charged H (charge equal to 0.75) to remove surface states and prevent charge transfer between both surfaces of the slab.

\section*{Supporting Figures}

\begin{figure}[ht!]{}
\begin{center}
\includegraphics[width=3.13in]{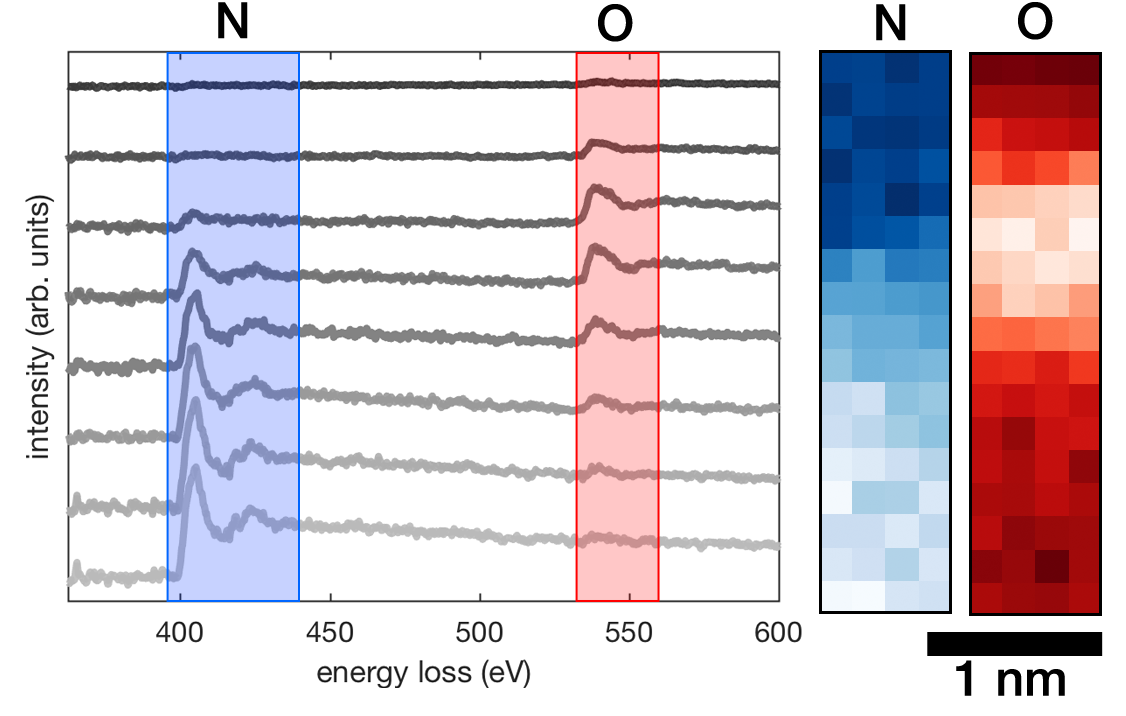}
\caption{Electron energy loss spectra for the GaN surface oxide. Colored boxes correspond to the integration areas used to produce the elemental distribution maps at the right.}
\label{fig:GaNeels}
\end{center}
\end{figure}

\begin{figure}[ht!]{}
\begin{center}
\includegraphics[width=3.13in]{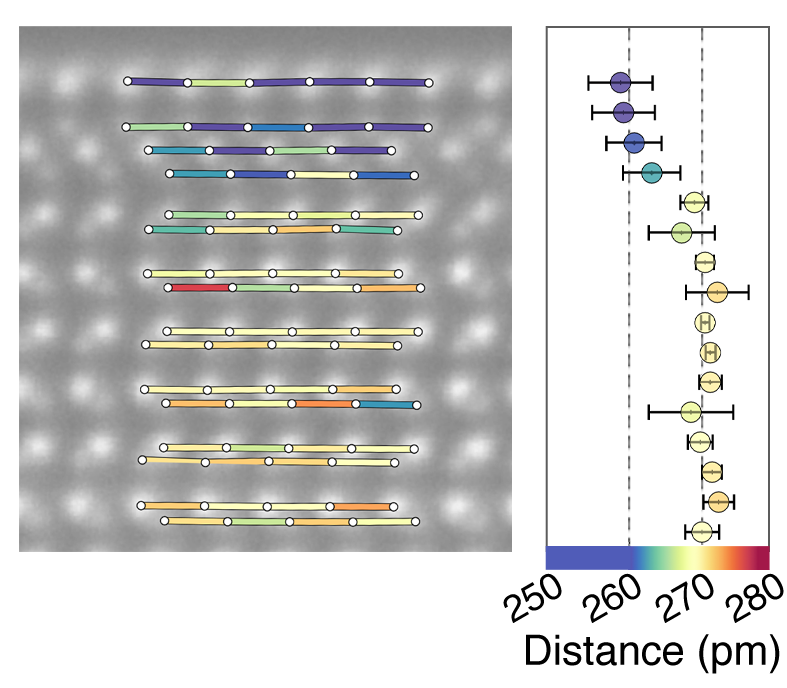}
\caption{(left) In--plane distance measurements from the AlN surface with color corresponding to the magnitude. (right) The average in-plane Al-Al distance for each layer}.
\label{fig:inplane}
\end{center}
\end{figure}

\begin{figure*}[ht!]{}
\begin{center}
\includegraphics[width=4in]{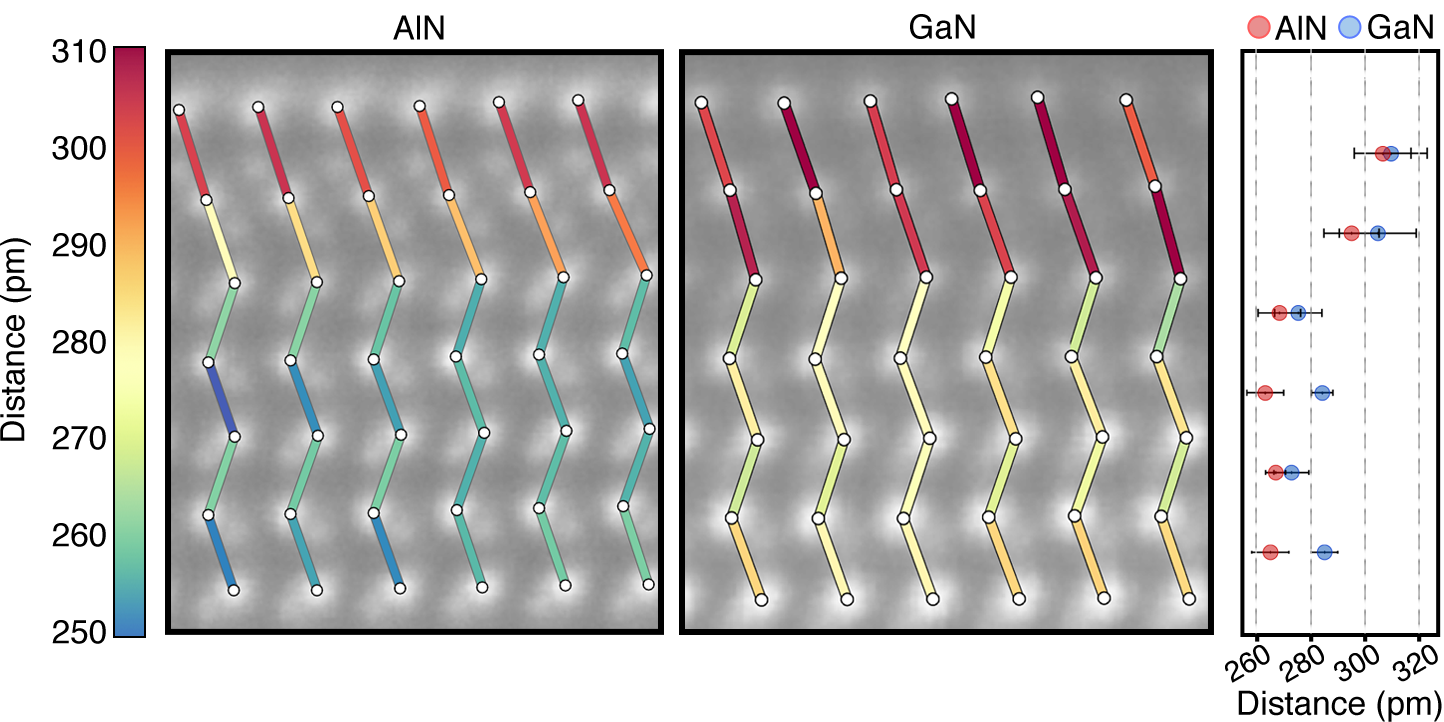}
 \caption{Distances between the group III positions viewed down the $\left[11\bar{2}0\right]$ axis for AlN and GaN. Due to the large difference in atomic number between Ga (Z=31) and N (Z=7), locating precise positions for the N atom columns in GaN was less reliable than for AlN. Therefore, rather than measuring Ga--N or Ga--O distances, Ga--Ga distances were measured and the relative changes for the surface with respect to the bulk were compared to the corresponding Al--Al distances in AlN.  For GaN and AlN, the Ga--Ga and Al--Al distances is 279 pm and 265 pm in the bulk, respectively. Within the surface oxide these distances increase to 307.5 pm  and 300.5 pm respectively.
 }
\label{fig:alnandgandist}
 \end{center}
 \end{figure*}

\begin{figure*}[ht!]{}
\begin{center}
\includegraphics[width=4in]{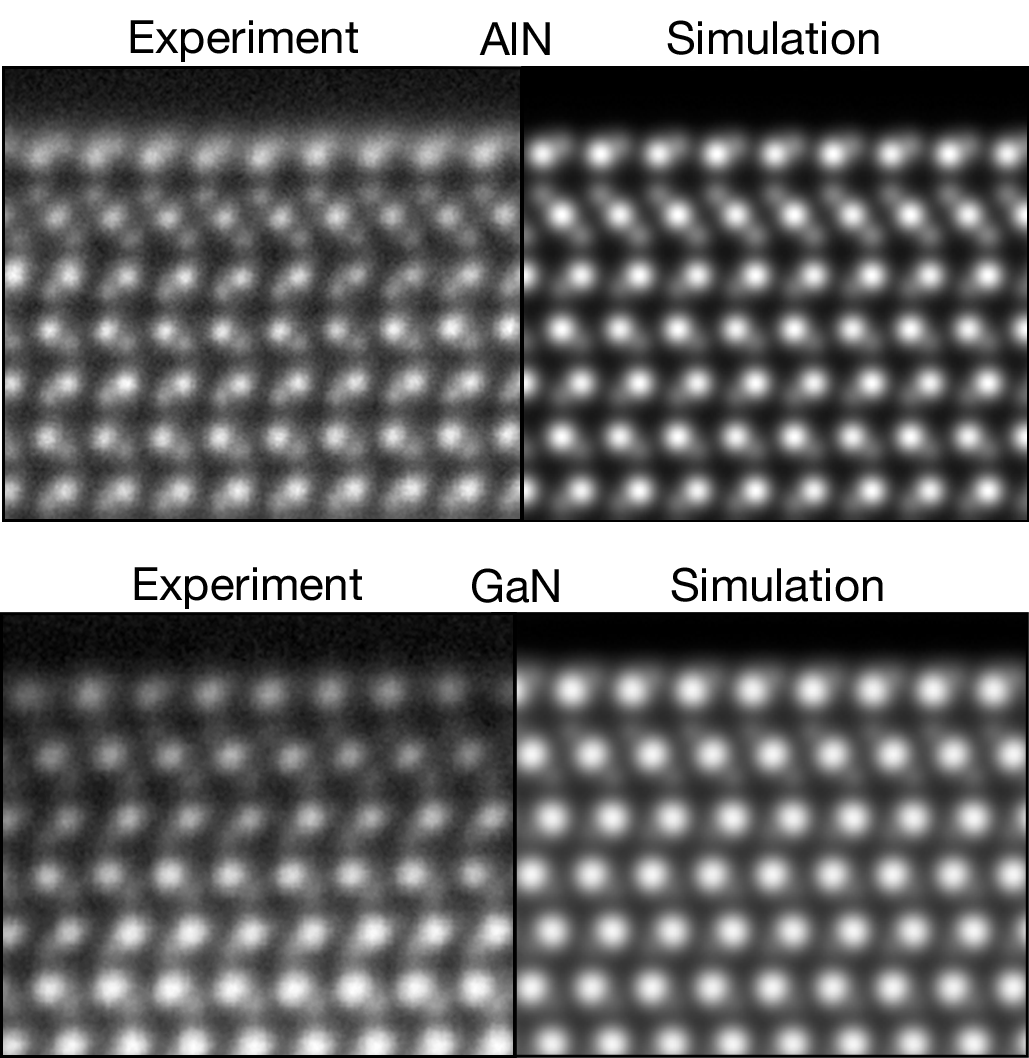}
\caption{Comparison of the experiment and multislice calculated images for the [11$\bar{2}$0] axis surfaces on AlN and GaN.}
\label{fig:supplSimulation}
\end{center}
\end{figure*}

\clearpage
\begin{table}[!ht]
    \centering
 \includegraphics[width=3.1in]{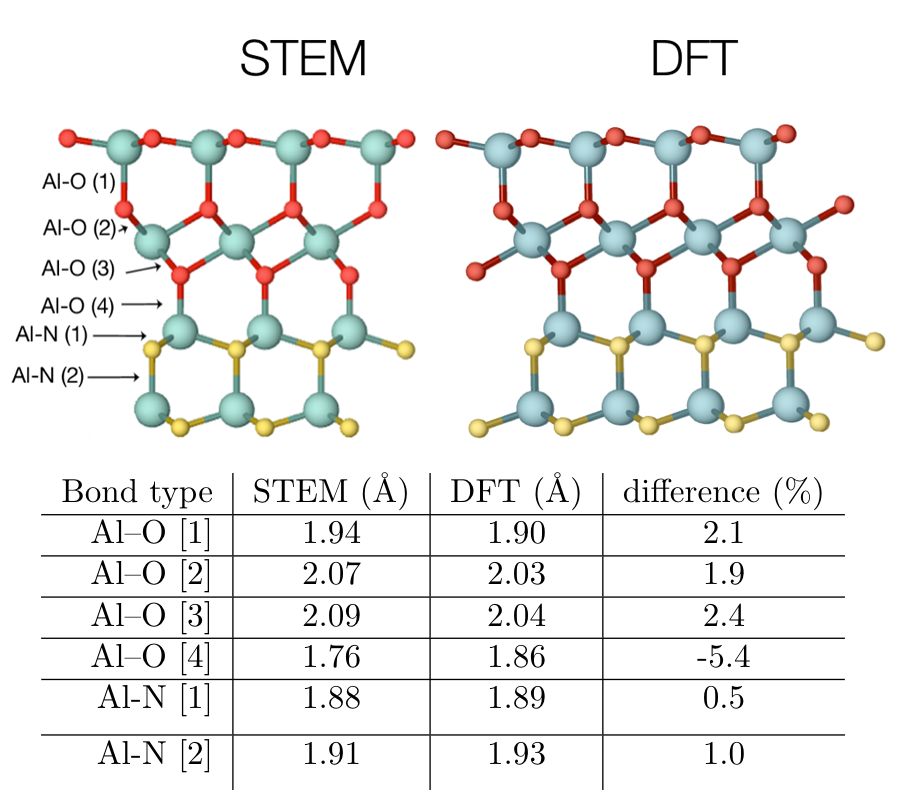}
 \caption{Differences in the bond lengths from the observed and simulated relaxed DFT structure STEM images.  The table shows the distance for each unique distance in the structure. STEM values are converted from the projected distance to the actual bond length.}
    \label{tab:STEMvsDFT}
\end{table}

\end{document}